\documentclass[eqsecnum,showkeys,showpacs,nofootinbib,aps,epsfig,color]{revtex4}
\makeatletter\@addtoreset{equation}{section}
\makeatother
\usepackage{amsmath}

\newbox\pippobox

\def \nn {\nonumber}
\def \p {\partial}

\begin{document}

\title{Quantum SUSY Algebra of $Q$-lumps in the Massive Grassmannian Sigma Model}

\author {Hiroaki Nakajima\footnote{E-mail address:
nakajima@skku.edu}, Phillial Oh\footnote{E-mail address:
ploh@skku.edu}, and Sunyoung Shin\footnote{E-mail address:
sihnsy@skku.edu}}

\affiliation{Department of Physics and Institute of Basic Science,
Sungkyunkwan University, Suwon 440-746 Korea}

\date{\today}
\begin{abstract}
We compute the $\mathcal{N}=2$ SUSY algebra of the massive
Grassmannian sigma model in 2+1 dimensions. We first rederive the
action of the model by using the Scherk-Schwarz dimensional
reduction from $\mathcal{N}=1$ theory in 3+1 dimensions. Then, we
perform the canonical quantization by using the Dirac method. We
find that a particular choice of the operator ordering yields the
quantum SUSY algebra of the $Q$-lumps with cental extension.
\end{abstract}
\pacs{11.30.Pb, 03.70+k, 11.10.-z}
 \keywords{$Q$-lump, supersymmetric algebra, BPS, Grassmannian model}

\maketitle
\section{introduction}
$Q$-lumps\cite{leese,abraham} are time dependent topological objects
which are stabilized by a Noether charge of global symmetry like
$Q$-balls\cite{coleman} as well as a topological charge. It is
well-known that $Q$-lumps are BPS objects and preserve a fraction of
supersymmetries. In this respect, there has been a great deal of
interest in this object. In particular, the $Q$-lump solutions in
massive sigma model are investigated in
\cite{gauntlett1,gauntlett2,Nitta,Nitta1,ward,bak-oh}, and the
relation to the D-brane configuration is also studied in
\cite{townsend1,townsend2,townsend3,townsend4}. In relation with
supersymmetric gauge theories, the massive sigma model can be
realized as an effective action of non-Abelian vortex strings, which
have been discovered recently \cite{tong,Nitta2,Nitta3}. $Q$-lumps
in supersymmetric gauge theories are examined in \cite{Nitta4}.

According to the well-known result of \cite{witten-olive,hlousek},
the SUSY algebra of the $Q$-lumps will include the central charges.
In relation with nonlinear sigma model, the central charges of the
${\mathrm{CP}}^N$ model were computed at the classical
level\cite{aoyama,rouhani}. In this paper, we explicitly compute the
quantum SUSY algebra of $Q$-lumps in the massive Grassmannian model.
The resulting SUSY algebra can be expected to change by a mass term
in the Hamiltonian compared with \cite{aoyama,rouhani}, but the
precise expressions of the central charges and the Hamiltonian
depend on the operator ordering and we find that enforcing the SUSY
algebra with central charges is closely tied with some particular
ordering prescription (See eqs. (\ref{U}), (\ref{qsusy}),
(\ref{Pzero})).

One way of obtaining the SUSY algebra is deriving it from
supersymmetric transformation rules. However, this passage usually
cannot deal with the operator ordering problem. Instead, we perform
the canonical quantization via the Dirac method by carefully taking
into account the ordering ambiguity. It turns out that a specific
choice of ordering yields the SUSY algebra with central extension.

We first derive the off-shell action of the massive Grassmannian
sigma model via Scherk-Schwarz dimensional
reduction\cite{scherk-schwarz}, then the Dirac analysis of
constraints is applied to get classical SUSY algebra. After that we
quantize the SUSY algebra by considering the ordering problem.

\section{$\mathcal{N}=2$ Off-Shell supersymmetric Grassmannian Action}
We consider the Grassmannian sigma model of which the target space
is the coset space $SU(N+M)/S(U(N)\times U(M)). %
$\footnote{Extended
supersymmetries of massive nonlinear sigma models have been studied
in various dimensions\cite{alvarez-gaume,gates,mass1,mass2,mass3,mass4,shifman1,shifman2,shifman3}.}%
~It is possible to obtain ${\mathcal{N}}=2$ supersymmetric action in
2+1 dimensions using superfield formalism by dimensional reduction
from ${\mathcal{N}}=1$ supersymmetric model in 3+1 dimensions. The
chiral and antichiral fields in the $(N+M)\times M$ matrix and the
vector fields in the $M\times M$ matrix in 3+1 dimensions are
defined as follows
\begin{eqnarray}
\Phi(x_L,\theta)&=&\phi(x_L)+\sqrt{2}\theta\psi(x_L)+{\theta\theta}F(x_L),~~~~~~~~~~
x_L^\mu=x^\mu-i\theta\sigma^\mu\bar{\theta},\\
\bar{\Phi}(x_R,\bar{\theta})&=&\phi(x_R)+\sqrt{2}\bar{\theta}\bar{\psi}(x_R)+\bar{\theta}\bar{\theta}\bar{F}(x_R),\,~~~~~~~~~
x_R^\mu=x^\mu+i\theta\sigma^\mu\bar{\theta},
\end{eqnarray}
\begin{eqnarray}
V=2\bar{\theta}\bar{\sigma}^\mu{\theta}A_\mu+i(\theta\theta)(\bar{\theta}\bar{\lambda})
-i(\bar{\theta}\bar{\theta})(\theta\lambda)+(\theta\theta)(\bar{\theta}\bar{\theta})\tau,
\end{eqnarray}
and the Lagrangian can be written in the form
\cite{wessbagger,aoyama}
\begin{eqnarray}
\int d^4\theta\,{\mathrm{tr}}[\bar{\Phi}\Phi e^V-V]
={\mathrm{tr}}\Big[\overline{D_\mu\phi}D^\mu\phi+\frac{i}{2}(-\overline{D_\mu\psi}\bar{\sigma}^\mu\psi+\bar{\psi}\bar{\sigma}^\mu
D_\mu\psi)+\bar{F}F-\frac{i}{\sqrt{2}}\bar{\psi}\phi\bar{\lambda}
+\frac{i}{\sqrt{2}}\bar{\phi}\psi\lambda+\tau(\bar{\phi}\phi-1)\Big]\label{n2lagrangian}.
\end{eqnarray}
In order to obtain massive model in 2+1 dimensions we apply the
Scherk-Schwarz dimensional reduction
\cite{scherk-schwarz,alvarez-gaume} specifying that the fields in
the $x^3$-direction are moving along orbits of the Killing vectors
$f(\phi)$ and $\bar{f}(\bar{\phi})$ in the Grassmannian manifold
\begin{eqnarray}
\frac{\p\phi}{\p{x}^3}=f(\phi),~~~\frac{\p\bar{\phi}}{\p{x}^3}=\bar{f}(\bar\phi),~~~\frac{\p\psi}{\p{x}^3}={\p}f(\phi)
\psi,~~~\frac{\p\bar{\psi}}{\p{x}^3}=\bar{\p}\bar{f}(\bar{\phi})\bar{\psi}.\label{x3}
\end{eqnarray}
The general forms of the Killing vector $f(\phi)$ and
$\bar{f}(\bar{\phi})$ are given by
\begin{eqnarray}
f(\phi)=i\mathcal{M}\phi,~~~~~~~~~~~~~~~~~~~~\bar{f}(\bar{\phi})=-i\bar{\phi}\mathcal{M},\label{Killing}
\end{eqnarray}
because the isometry $SU(N+M)$ is linearly realized and the matrix
$\mathcal{M}$ is diagonal element of it%
\footnote{However, the traceless condition for $\mathcal{M}$ can be
relaxed by the constraint \eqref{const1} since the Lagrangian is
invariant under the constant shift $\mathcal{M}$ to
$\mathcal{M}+cI$. Due to this fact, some of $\mathcal{M}$'s can be
shifted to projection matrices, which is used in (\ref{action2})}.
We substitute (\ref{x3}) and (\ref{Killing}) into
(\ref{n2lagrangian}) to obtain (with $A_3\equiv\sigma$)
\begin{eqnarray}
{\mathcal{S}}&=&\int
d^3x\,\mathrm{tr}\Big[|D_\mu\phi|^2+\frac{i}{2}(-\overline{D_\mu\psi}\gamma^\mu\psi
+\bar{\psi}\gamma^\mu D_\mu\psi)
+\bar{F}F-\bar{\phi}\mathcal{M}^2\phi+2\bar{\phi}\mathcal{M}\phi\sigma-\bar{\phi}\phi\sigma^2+\bar{\psi}\mathcal{M}\psi
-\bar{\psi}\psi\sigma\nn\\
&&~~~~~~~~~~-\frac{i}{\sqrt{2}}\bar{\psi}\phi\bar{\lambda}+\frac{i}{\sqrt{2}}\bar{\phi}\psi\lambda+
\tau(\bar{\phi}\phi-1)\Big].
\end{eqnarray}
By constraints of the system
\begin{eqnarray}
\bar{\phi}\phi=I\,,~~~~~\bar{\phi}\psi_\alpha=0=\bar{\psi}_\alpha\phi,
\label{const1}
\end{eqnarray}
and eliminating auxiliary fields
\begin{alignat}{3}
F&=0, &
\bar{F}&=0,\notag\\
\sigma&=\frac{1}{2}(2\bar{\phi}\mathcal{M}\phi-\bar{\psi}\psi),
&\qquad A^\mu
&=\frac{1}{2}(i\p^\mu\bar{\phi}\phi-i\bar{\phi}\p^\mu\phi-\bar{\psi}\gamma^\mu\psi).
\end{alignat}
we get the action
\begin{eqnarray}
{\mathcal{S}}&=&\int
d^3x\,\mathrm{tr}\Big[|\p_\mu\phi|^2+\frac{i}{2}(-\p_\mu\bar{\psi}\gamma^\mu\psi+\bar{\psi}\gamma^\mu\p_\mu\psi)
-\frac{1}{4}(i\p_\mu\bar{\phi}\phi-i\bar{\phi}\p_\mu\phi-\bar{\psi}\gamma_\mu\psi)^2\\
&&~~~~~~~~~~+(\bar{\phi}\mathcal{M}\phi-\frac{1}{2}\bar{\psi}\psi)^2
-\bar{\phi}\mathcal{M}^2\phi+\bar{\psi}\mathcal{M}\psi\Big].
\label{onshell}
\end{eqnarray}
With the definition of $\mathcal{M}=m\mathcal{P}$ where
$\mathcal{P}$ is the $(N+M)\times(N+M)$ Hermitian projection matrix
satisfying $\mathcal{P}^2=\mathcal{P}$ and $m$ is a real positive
number, the action is
\begin{eqnarray}
{\mathcal{S}}&=&\int
d^3x\,\mathrm{tr}\Big[|\p_\mu\phi|^2+\frac{i}{2}(-\p_\mu\bar{\psi}\gamma^\mu\psi+\bar{\psi}\gamma^\mu\p_\mu\psi)
-\frac{1}{4}(i\p_\mu\bar{\phi}\phi-i\bar{\phi}\p_\mu\phi-\bar{\psi}\gamma_\mu\psi)^2\nn\\
&&~~~~~~~~~~+(m\bar{\phi}\mathcal{P}\phi
-\frac{1}{2}\bar{\psi}\psi)^2-m^2\bar{\phi}\mathcal{P}\phi+m\bar{\psi}\mathcal{P}\psi\Big],
\label{action2}
\end{eqnarray}
which is the same as the one given in \cite{bak-oh}.
\section{Dirac Analysis}
In this section, we perform the Dirac analysis, which is useful to
calculate the algebra of constrained system, to obtain the SUSY
algebra of the action (\ref{action2}). The massless supersymmetric
${\mathrm{CP}}^N$ model was studied in \cite{aoyama,rouhani}. The
Hamiltonian of the system is
\begin{eqnarray}
H&=&\int d^2x \,\mathrm{tr}\Big[
\Pi\bar{\Pi}-|\partial_i\phi|^2+\frac{1}{4}|\partial_i\bar{\phi}\phi-\bar{\phi}\partial_i\phi|^2+\frac{i}{2}(\partial_i\bar{\psi}\gamma^i\psi-
\bar{\psi}\gamma^i\partial_i\psi)
-\frac{i}{2}(\bar{\psi}\gamma^i\psi)(\partial_i\bar{\phi}\phi-\bar{\phi}\partial_i\phi)\nn\\
&&~~~~~~~~~~+m^2\{(\bar{\phi}\mathcal{P}\phi)-(\bar{\phi}\mathcal{P}\phi)^2\}
+m(\bar{\phi}\mathcal{P}\phi)(\bar{\psi}\psi)-m(\bar{\psi}\mathcal{P}\psi)+\frac{1}{4}(\bar{\psi}\gamma_i\psi)
(\bar{\psi}\gamma^i\psi)-\frac{1}{4}(\bar{\psi}\psi)^2\Big],\label{hamiltonian0}
\end{eqnarray}
where the conjugate momenta are given by
\begin{eqnarray}
\Pi_a^{~i}=\frac{\delta}{\delta\dot{\phi}_i^{~a}}{\int
d^2x\mathcal{L}}=\overline{(D_0\phi)}_a^{~i},~~~~
\bar{\Pi}_i^{~a}=\frac{\delta}{\delta\dot{\bar{\phi}}_a^{~i}}{\int
d^2x\mathcal{L}}=(D_0\phi)_i^{~a}.
\end{eqnarray}
We use Poisson brackets defined as follows
\begin{eqnarray}
\Big\{\phi_i^{~a}(x),\Pi_b^{~j}(y)\Big\}_{\scriptscriptstyle\mathrm{P.B.}}&=&\delta_b^{~a}\delta_i^{~j}\delta(x-y),\nn\\
\Big\{\bar{\phi}_a^{~i}(x),\bar{\Pi}_j^{~b}(y)\Big\}_{\scriptscriptstyle\mathrm{P.B.}}&=&\delta_a^{~b}\delta_j^{~i}\delta(x-y),\nn\\
\Big\{\psi_{\alpha i}^{~a}(x),\psi^{\dagger\beta}
{_b^{~j}}(y)\Big\}_{\scriptscriptstyle\mathrm{P.B.}}&=&-i\delta_\alpha^{~\beta}\delta_b^{~a}\delta_i^{~j}\delta(x-y).
\end{eqnarray}
There are one Gauss law constraint
\begin{eqnarray}
\bar{\phi}\bar{\Pi}-\Pi\phi-i\bar{\psi}\gamma^0\psi=0,
\end{eqnarray}
and four second class constraints
\begin{eqnarray}
&&C^1{_a^{~b}}=\bar{\phi}_a^{~i}\phi_i^{~b}-\delta_a^{~b}\approx0,~~~~~~~
C^2{_a^{~b}}=\Pi_a^{~i}\phi_i^{~b}+\bar{\phi}_a^{~i}\bar{\Pi}_i^{~b}\approx0,
\nn\\
&&C^3{_a^{~b}}=\psi^\dagger{_a^{~i}}\phi_i^{~b}\approx0,~~~~~~~~~~~~
C^4{_a^{~b}}=\bar{\phi}_a^{~i}\psi_i^{~b}\approx0.
\label{secondconstraint}
\end{eqnarray}
We label the second class constraints as $C_A\equiv(C^1{_a^{~b}},
C^2{^a_{~b}}, C^3{^a_{~b}}, C^4{_a^{~b}})~(A=1,2,\ldots, 4M^2)$, and
then the Dirac matrix is given by
\begin{eqnarray}
\Omega=\left\{C_A,C_B\right\}_{\scriptscriptstyle\mathrm{P.B.}}
= \left[\begin{array}{cccc} 0&X&0&0\\ -X^\mathrm{T}&Y&0&0\\0&0&0&Z\\
0&0&Z^\mathrm{T}&0\end{array}\right],~~~~~~
\Omega^{-1}= \left[\begin{array}{cccc} X^{\mathrm{T}-1}YX^{-1}&X&0&0\\ -X^{\mathrm{T}-1}&0&0&0\\0&0&0&Z^{\mathrm{T}-1}\\
0&0&Z^{-1}&0\end{array}\right],
\end{eqnarray}
where
\begin{eqnarray}
X_{AB}&\equiv&X{_a^{~b}};{_c^{~d}}=\left\{C^1{_a^{~b}},
C^2{_c^{~d}}\right\}_{\scriptscriptstyle\mathrm{P.B.}}=2\delta_a^{~d}\,\delta_c^{~b},\nn\\
Y_{AB}&\equiv&\,Y{_a^{~b}};{_c^{~d}}=\left\{C^2{_a^{~b}},
C^2{_c^{~d}}\right\}_{\scriptscriptstyle\mathrm{P.B.}}=
\delta_a^{~d}\,(\Pi\phi-\bar{\phi}\bar{\Pi})_c^{~b}-\delta_c^{~b}(\Pi\phi-\bar{\phi}\bar{\Pi})_a^{~d},\nn\\
Z_{AB}&\equiv&\,Z{{_a^{~b}}\,;{_c^{~d}}}=\left\{C^3{_a^{~b}},C^4{_c^{~d}}\right\}_{\scriptscriptstyle\mathrm{P.B.}}
=-i\delta_a^{~d}\,\delta_c^{~d}.
\end{eqnarray}

The Dirac bracket is defined by
\begin{eqnarray}
\Big\{P_A(x),Q_B(y)\Big\}_{\scriptscriptstyle\mathrm{D}}=
\Big\{P_A(x),Q_B(y)\Big\}_{\scriptscriptstyle\mathrm{P.B.}}-\int{dzdz'}\Big\{P_A(x),C_E(z)\Big\}_{\scriptscriptstyle\mathrm{P.B.}}
\Omega^{-1EF}(z,z')\Big\{C_F(z'),Q_B(y)\Big\}_{\scriptscriptstyle\mathrm{P.B.}}.
\end{eqnarray}
Then the Dirac brackets between the physical variables are given by
\begin{eqnarray}
\Big\{\phi_i^{~a}(x),\phi_j^{~b}(y)\Big\}_{\scriptscriptstyle\mathrm{D}}&=&0,\\
\Big\{\bar{\phi}_a^{~\,i}(x),\bar{\phi}_b^{~\,j}(y)\Big\}_{\scriptscriptstyle\mathrm{D}}&=&0,\\
\Big\{\phi_i^{~a}(x),\bar{\phi}_b^{~\,j}(y)\Big\}_{\scriptscriptstyle\mathrm{D}}&=&0,\\
\Big\{\phi_i^{~a}(x),\Pi_b^{~j}(y)\Big\}_{\scriptscriptstyle\mathrm{D}}&=&\delta_b^{~a}\left(\delta_i^{~j}-
\frac{1}{2}\phi_i^{~c}\bar{\phi}_c^{~j}\right)\delta(x-y),
\label{var1}\\
\Big\{\bar{\phi}_a^{~\,i}(x),\bar{\Pi}_j^{~b}(y)\Big\}_{\scriptscriptstyle\mathrm{D}}&=&
\delta_a^{~b}\left(\delta_j^{~i}-\frac{1}{2}\phi_j^{~c}\bar{\phi}_c^{~i}\right)\delta(x-y),\\
\Big\{\phi_i^{~a}(x),\bar{\Pi}_j^{~b}(y)\Big\}_{\scriptscriptstyle\mathrm{D}}&=&
-\frac{1}{2}\phi_i^{~b}\phi_j^{~a}\delta(x-y),\\
\Big\{\bar{\phi}_a^{~i}(x),\Pi_b^{~j}(y)\Big\}_{\scriptscriptstyle\mathrm{D}}&=&
-\frac{1}{2}\bar{\phi}_b^{~i}\bar{\phi}_a^{~j}\delta(x-y),
\end{eqnarray}
\begin{eqnarray}
\Big\{\Pi_a^{~i}(x),\Pi_b^{~j}(y)\Big\}_{\scriptscriptstyle\mathrm{D}}&=&
\Big[\,\frac{1}{4}\bar{\phi}_b^{~i}\bar{\phi}_c^{~j}\left(\Pi\phi-\bar{\phi}\bar{\Pi}\right)_a^{~c}
-\frac{1}{4}\bar{\phi}_c^{~i}\bar{\phi}_a^{~j}\left(\Pi\phi-\bar{\phi}\bar{\Pi}\right)_b^{~c}
-\frac{1}{2}\left(\bar{\phi}_b^{~i}\Pi_a^{~j}-\Pi_b^{~i}\bar{\phi}_a^{~j}\right)\Big]\delta(x-y),~~\\
\Big\{\bar{\Pi}_i^{~a}(x),\bar{\Pi}_j^{~b}(y)\Big\}_{\scriptscriptstyle\mathrm{D}}&=&
\Big[\,\frac{1}{4}\phi_i^{~c}\phi_j^{~a}\left(\Pi\phi-\bar{\phi}\bar{\Pi}\right)_c^{~b}
-\frac{1}{4}\phi_i^{~b}\phi_j^{~c}\left(\Pi\phi-\bar{\phi}\bar{\Pi}\right)_c^{~a}
-\frac{1}{2}\left(\phi_i^{~b}\bar{\Pi}_j^{~a}-\bar{\Pi}_i^{~b}\phi_j^{~a}\right)\Big]\delta(x-y),~~~~\\
\Big\{\Pi_a^{~i}(x),\bar{\Pi}_j^{~b}(y)\Big\}_{\scriptscriptstyle\mathrm{D}}&=&
\Big[\,\frac{1}{4}\bar{\phi}_c^{~i}\phi_j^{~c}\left(\Pi\phi-\bar{\phi}\bar{\Pi}\right)_a^{~b}-
\frac{1}{4}\bar{\phi}_c^{~i}\phi_j^{~d}\left(\Pi\phi-\bar{\phi}\bar{\Pi}\right)_d^{~c}\delta_a^{~b}
-\frac{1}{2}\left(\bar{\phi}_c^{~i}\bar{\Pi}_j^{~c}-\Pi_c^{~i}\phi_j^{~c}\right)\delta_a^{~b}\hspace{7mm}\nn\\
&&+i\psi^\dagger{_c^{~i}}\psi_j^{~c}\delta_a^{~b}\Big]\delta(x-y),\\
\Big\{\Pi_a^{~i}(x),{\psi^{\dagger\alpha}}{_b^{~j}}(y)\Big\}_{\scriptscriptstyle\mathrm{D}}&=&
\psi^{\dagger \alpha}{_b^{~i}}\bar{\phi}_a^{~j}\delta(x-y),\\
\Big\{\bar{\Pi}_i^{~a}(x),\psi_\alpha{_j^{~b}}(y)\Big\}_{\scriptscriptstyle\mathrm{D}}&=&
\psi_{\alpha i}^{~~b}\phi_j^{~a}\delta(x-y),\\
\Big\{\psi_{\alpha
i}^{~~a}(x),\psi^{\dagger\beta}{_b^{~j}}(y)\Big\}_{\scriptscriptstyle\mathrm{D}}&=&
-i\delta_\alpha^{~\beta}\delta_b^{~a}\left(\delta_i^{~j}-\phi_i^{~c}\bar{\phi}_c^{~j}\right)\delta(x-y)\,.\label{var2}
\end{eqnarray}

We use Noether procedure to obtain the various conserved charges.
First the supercharge is
\begin{eqnarray}
Q_\alpha=\int d^2 x
\,\mathrm{tr}\Big[\Pi\psi_\alpha+\partial_i\bar{\phi}(\gamma^i\gamma^0\psi)_\alpha-
im\bar{\phi}\mathcal{P}(\gamma^0\psi)_\alpha\Big].
\label{supercharge}
\end{eqnarray}
The Hamiltonian is given by (\ref{hamiltonian0}) and the momenta
\begin{eqnarray}
P^i=\int d^2x \, \mathrm{tr}\Big[
\Pi\partial^i\phi+\partial^i\bar{\phi}\bar{\Pi}+\frac{i}{2}(\bar{\psi}\gamma^0\p^i\psi-\p^i\bar{\psi}\gamma^0\psi)
+\frac{1}{2}(-i\Pi\phi+i\bar{\phi}\bar{\Pi}+\bar{\psi}\gamma^0\psi)(i\p^i\bar{\phi}\phi-i\bar{\phi}\p^i\phi-
\bar{\psi}\gamma^i\psi)\Big],\hspace{2mm} \label{momenta}
\end{eqnarray}
where the last term in (\ref{momenta}) is a gauge degree of freedom.
There is also a symmetry under the transformations
$\delta\phi=i\mathcal{P}\phi$ and $\delta\psi=i\mathcal{P}\psi$, and
the corresponding scalar Noether charge is
\begin{eqnarray}
U=\int d^2 x \,\mathrm{tr}\Big[i\Pi\mathcal{P}\phi-i\bar{\phi}
\mathcal{P}\bar{\Pi}-\bar{\psi}\mathcal{P}\gamma^0\psi\Big].\label{U}
\end{eqnarray}

We compute explicitly to obtain the Dirac brackets among the
supercharges using the relations (\ref{var1})-(\ref{var2})
\begin{eqnarray}
\Big\{Q_\alpha,Q^{\dagger\beta}\Big\}_{\scriptscriptstyle\mathrm{D}}
=-i(\gamma^\mu\gamma^0)_\alpha^{~\beta}P_\mu-im\gamma^0{_\alpha^{~\beta}}U-i\gamma^0{_\alpha^{~\beta}}(2{\pi}T)
+\gamma^i{_\alpha^{~\beta}}R_i\,,\label{diracsuper}
\end{eqnarray}
where
\begin{eqnarray}
T&=&\frac{i}{2\pi}\int d^2x
\,\mathrm{tr}\Big[\epsilon^{ij}\{(\partial_i\bar{\phi})(\partial_j\phi)+\frac{i}{2}\partial_i(\bar{\psi}\gamma_j\psi)
\}\Big],\label{T}\\
R_i&=&\int
d^2x\,\mathrm{tr}\Big[\frac{1}{2}\p_i(\bar{\psi}\psi)+m\p_i(\bar{\phi}\mathcal{P}\phi)\Big].\label{Y}
\end{eqnarray}
\section{Quantization of Dirac brackets}
We quantize Dirac brackets (\ref{var1})-(\ref{var2}). Assuming that
the ordering of the second class constraints are fixed as in
(\ref{secondconstraint}), one of the possible choices of the
ordering which makes all the dynamical variables commute with the
second class constraints is given by
\begin{eqnarray}
\Big[\phi_i^{~a}(x),\Pi_b^{~j}(y)\Big]&=&i\delta_b^{~a}\left(\delta_i^{~j}-\frac{1}{2}\phi_i^{~c}\bar{\phi}_c^{~j}\right)\delta(x-y),\\
\Big[\bar{\phi}_a^{~i}(x),\bar{\Pi}_j^{~b}(y)\Big]&=&i\delta_a^{~b}\left(\delta_j^{~i}-
\frac{1}{2}\phi_j^{~c}\bar{\phi}_c^{~i}\right)\delta(x-y),\\
\Big[\phi_i^{~a}(x),\bar{\Pi}_j^{~b}(y)\Big]&=&-\frac{i}{2}\phi_i^{~b}\phi_j^{~a}\delta(x-y),\\
\Big[\bar{\phi}_a^{~i}(x),\Pi_b^{~j}(y)\Big]&=&-\frac{i}{2}\bar{\phi}_b^{~i}\bar{\phi}_a^{~j}\delta(x-y),\\
\Big[\Pi_a^{~i}(x),\Pi_b^{~j}(y)\Big]&=&
\Big[\frac{i}{4}\left(\Pi\phi-\bar{\phi}\bar{\Pi}\right)_a^{~c}\bar{\phi}_b^{~i}\bar{\phi}_c^{~j}
-\frac{i}{4}\left(\Pi\phi-\bar{\phi}\bar{\Pi}\right)_b^{~c}\bar{\phi}_c^{~i}\bar{\phi}_a^{~j}
-\frac{i}{2}\left(\bar{\phi}_b^{~i}\Pi_a^{~j}-\bar{\phi}_a^{~j}\Pi_b^{~i}\right)\Big]\delta(x-y),\label{PiPi}\\
\Big[\bar{\Pi}_i^{~a}(x),\bar{\Pi}_j^{~b}(y)\Big]&=&
\Big[\frac{i}{4}\phi_i^{~c}\phi_j^{~a}\left(\Pi\phi-\bar{\phi}\bar{\Pi}\right)_c^{~b}
-\frac{i}{4}\phi_i^{~b}\phi_j^{~c}\left(\Pi\phi-\bar{\phi}\bar{\Pi}\right)_c^{~a}
-\frac{i}{2}\left(\bar{\Pi}_j^{~a}\phi_i^{~b}-\bar{\Pi}_i^{~b}\phi_j^{~a}\right)\Big]\delta(x-y),
\label{barPibarPi}
\end{eqnarray}
\begin{eqnarray}
\Big[\Pi_a^{~i}(x),\bar{\Pi}_j^{~b}(y)\Big]&=&
\Big[\frac{i}{4}\bar{\phi}_c^{~i}\phi_j^{~c}\left(\Pi\phi-\bar{\phi}\bar{\Pi}\right)_a^{~b}-
\frac{i}{4}\bar{\phi}_c^{~i}\phi_j^{~d}\left(\Pi\phi-\bar{\phi}\bar{\Pi}\right)_d^{~c}\delta_a^{~b}
-\frac{i}{2}\left(\bar{\phi}_c^{~i}\bar{\Pi}_j^{~c}-\phi_j^{~c}\Pi_c^{~i}\right)\delta_a^{~b}\hspace{7mm}\nn\\
&&-\psi^\dagger{_c^{~i}}\psi_j^{~c}\delta_a^{~b}-h\left\{\psi_j^{~c},\psi^\dagger{_c^{~i}}\right\}\delta_a^{~b}\Big]\delta(x-y),\label{PibarPi}\\
\Big[\Pi_a^{~i}(x),{\psi^{\dagger\alpha}}{_b^{~j}}(y)\Big]&=&
i\psi^{\dagger \alpha}{_b^{~i}}\bar{\phi}_a^{~j}\delta(x-y),\\
\Big[\bar{\Pi}_i^{~a}(x),\psi_\alpha{_j^{~b}}(y)\Big]&=&i\psi_{\alpha i}^{~~b}\phi_j^{~a}\delta(x-y),\\
\Big\{\psi_{\alpha
i}^{~~a}(x),\psi^{\dagger\beta}{_b^{~j}}(y)\Big\}&=&
\delta_\alpha^{~\beta}\delta_b^{~a}\left(\delta_i^{~j}-\phi_i^{~c}\bar{\phi}_c^{~j}\right)\delta(x-y)\,.
\end{eqnarray}
Note that with the above choice, (\ref{PiPi}) and (\ref{barPibarPi})
vanish for identical indices. This is the same as the method of
\cite{Han}, where the Dirac analysis is used for bosonic
${\mathrm{CP}}^N$ model. In the above (\ref{PibarPi}) the ordering
parameter $h$ is undetermined. It will be fixed by the SUSY algebra.

Since the supercharge in (\ref{supercharge}) does not have any
ordering ambiguity, a straightforward computation yields the
following quantum SUSY algebra.
\begin{eqnarray}
\Big\{Q_\alpha,Q^{\dagger\beta}\Big\} &=&\delta_\alpha^{~\beta}\int
d^2x \,\mathrm{tr} \Big[\Pi\bar{\Pi}-|\partial_i\bar{\phi}|^2+
\frac{1}{4}|\partial_i\bar{\phi}\phi-\bar{\phi}\partial_i\phi|^2+\frac{i}{2}\left(\partial_i\bar{\psi}\gamma^i\psi-
\bar{\psi}\gamma^i\partial_i\psi\right)
-\frac{i}{2}\left(\bar{\psi}\gamma^i\psi\right)\left(\partial_i\bar{\phi}\phi-\bar{\phi}\partial_i\phi\right)\nn\\
&&\hspace{1.8cm}+m^2\{\left(\bar{\phi}\mathcal{P}\phi)-(\bar{\phi}\mathcal{P}\phi\right)^2\}
+m\left(\bar{\phi}\mathcal{P}\phi\right)\left(\bar{\psi}\psi\right)-m\left(\bar{\psi}\mathcal{P}\psi\right)+
\frac{1}{4}\left(\bar{\psi}\gamma_i\psi\right)
\left(\bar{\psi}\gamma^i\psi\right)-\frac{1}{4}\left(\bar{\psi}\psi\right)^2\Big]\nn\\
&&+\int d^2x ~\delta_\alpha^{~\beta}
\frac{1}{2}\delta_a^{~a}(\bar{\psi}\gamma^0\psi)_b^{~b}
+(2h+\frac{1}{2})\delta_a^{~a}\left\{(\bar{\psi}\gamma^0)^{\beta}\psi_\alpha\right\}_b^{~b}\nn\\
&&+(\gamma^i\gamma^0)_\alpha^{~\beta}\int d^2x \,\mathrm{tr}
\Big[\Pi\p_i\phi+\p_i\bar{\phi}\bar{\Pi}+\frac{i}{2}(\bar{\psi}\gamma^0\p_i\psi-\p_i\bar{\psi}\gamma^0\psi)\Big]\nn\\
&&+m\gamma^0{_\alpha^{~\beta}}\int d^2x \,\mathrm{tr}
\Big[i\Pi\mathcal{P}\phi-i\bar{\phi}
\mathcal{P}\bar{\Pi}-\bar{\psi}\mathcal{P}\gamma^0\psi\Big]\nn\\
&&+\gamma^0{_\alpha^{~\beta}}(2\pi)\int d^2x
\,\mathrm{tr}\Big[\frac{i\epsilon^{ij}}{2\pi}\{(\partial_i\bar{\phi})(\partial_j\phi)+\frac{i}{2}\partial_i(\bar{\psi}\gamma_j\psi)
\}\Big]\nn\\
&&+i\gamma^i{_\alpha^{~\beta}}\int
d^2x\,\mathrm{tr}\Big[\frac{1}{2}\p_i(\bar{\psi}\psi)+m\p_i(\bar{\phi}\mathcal{P}\phi)\Big].
\end{eqnarray}
The first term in the third line arises from the ordering of the
last two quartic terms of $\psi$ in (\ref{hamiltonian0}), and
therefore it can be absorbed in the definition of energy. We may
appropriately fix the parameter $h$ to eliminate the second term in
the line to make sure that the SUSY algebra closes at quantum level.
We choose $h=-\frac{1}{4}$ to get quantum SUSY algebra of the form,
\begin{eqnarray}
\Big\{Q_\alpha,Q^{\dagger\beta}\Big\}
=(\gamma^\mu\gamma^0)_\alpha^{~\beta}P_\mu+m\gamma^0{_\alpha^{~\beta}}U+\gamma^0{_\alpha^{~\beta}}(2{\pi}T)
+i\gamma^i{_\alpha^{~\beta}}R_i,\label{qsusy}
\end{eqnarray}
where the quantum Hamiltonian is given by
\begin{eqnarray}
P^0=\int d^2x\, \mathrm{tr}\Big[
\Pi\bar{\Pi}-|\partial_i\bar{\phi}|^2+\frac{1}{4}|\partial_i\bar{\phi}\phi-\bar{\phi}\partial_i\phi|^2+\frac{i}{2}(\partial_i\bar{\psi}\gamma^i\psi-
\bar{\psi}\gamma^i\partial_i\psi)
-\frac{i}{2}(\bar{\psi}\gamma^i\psi)(\partial_i\bar{\phi}\phi-\bar{\phi}\partial_i\phi)~~~~~~~~~~~~\nn\\
~~~~~~~+m^2\{(\bar{\phi}\mathcal{P}\phi)-(\bar{\phi}\mathcal{P}\phi)^2\}
+m(\bar{\phi}\mathcal{P}\phi)(\bar{\psi}\psi)-m(\bar{\psi}\mathcal{P}\psi)+\frac{1}{4}(\bar{\psi}\gamma_i\psi)
(\bar{\psi}\gamma^i\psi)-\frac{1}{4}(\bar{\psi}\psi)^2+\frac{1}{2}M(\bar{\psi}\gamma^0\psi)\Big].\,\label{Pzero}
\end{eqnarray}
Here $M$ is the number of color indices and the other operators are
the same as (\ref{momenta}), (\ref{U}), (\ref{T}) and (\ref{Y}).

It can be shown that the SUSY algebra (\ref{qsusy}) can be rewritten
as
\begin{eqnarray}
\Big\{Q_{\pm\alpha},Q_\pm^{\,\dagger\,\beta}\Big\}
&=&\frac{1}{4}\Big[\{\gamma^0,\gamma^\mu\}_\alpha^{~\beta}\pm(\gamma^\mu+\gamma^{\mu\dagger})_\alpha^{~\beta}\Big]P_\mu
+\frac{1}{2}m(\gamma^0{_\alpha^{~\beta}}\pm\delta_\alpha^{~\beta})U
+\frac{1}{2}(\gamma^0{_\alpha^{~\beta}}\pm\delta_\alpha^{~\beta})2\pi{T},\label{quanSUSYalg}\\
\Big\{Q_{\pm\alpha},Q_\pm^{\,\dagger\,\alpha}\Big\}&=&P_0\pm mU\pm 2\pi T, \label{qsusybps}\\
\Big\{Q_{\pm\alpha},Q_\mp^{\,\dagger\,\beta}\Big\}&=&\frac{1}{4}\Big\{[\gamma^\mu,\gamma^0]_\alpha^{~\beta}\mp
(\gamma^\mu-\gamma^{\mu\dagger})_\alpha^{~\beta}\Big\}P_\mu+\frac{i}{2}(\gamma^i{_\alpha^{~\beta}}
\pm(\gamma^0\gamma^i){_\alpha^{~\beta}})R_i,\\
\Big\{Q_{\pm\alpha},Q_\mp^{\,\dagger\,\alpha}\Big\}&=&0,
\end{eqnarray}
where we redefine supercharges as
$Q_{\pm\alpha}\equiv\left(\frac{1\pm\gamma^0}{2}Q\right)_\alpha$.
The explicit forms of the $Q_\pm$ and $Q_\pm^\dagger$ are
\begin{eqnarray}
Q_\pm&=&\frac{1}{2}\int
d^2x\,\mbox{tr}\Big[(1\pm\gamma^0)(\p_0\bar{\phi}\mp
im\bar{\phi}{\mathcal{P}})\psi+(\gamma^i\gamma^0\mp\gamma^i)\p_i\bar{\phi}\psi\Big],\\
Q_\pm^\dagger&=&\frac{1}{2}\int
d^2x\,\mbox{tr}\Big[\psi^\dagger(1\pm\gamma^0)(\p_0\phi\pm
im{\mathcal{P}}\phi)+\psi^\dagger(\gamma^i\gamma^0\pm\gamma^i)\p_i\phi\Big].
\end{eqnarray}
From (\ref{qsusybps}), the energy is bounded as $P_0\geq m|U|+2\pi
|T|$ and the saturation occurs when $Q_+=0$ or $Q_-=0$, i.e. the
following Bogomol¡¯nyi equations are satisfied
\begin{eqnarray}
\p_0\phi{\pm}im{\mathcal{P}}\phi&=&0,\\
\p_i\phi{\mp}\,i\epsilon_{ij}\p^j\phi&=&0,~~(\epsilon_{12}=1),
\end{eqnarray}
which shows that the $Q$-lumps are $\frac{1}{2}$BPS objects
\cite{bak-oh}. With these BPS equations satisfied, the energy is
given by
\begin{eqnarray}
P_0=m|U|+2\pi |T|.
\end{eqnarray}
\section{Conclusion}
In this paper, we studied $\mathcal{N}=2$ massive Grassmannian sigma
model in 2+1 dimensions. We derived the off-shell action by
Scherk-Schwarz dimensional reduction from $\mathcal{N}=1$ formalism
in 3+1 dimensions. We performed canonical analysis via Dirac method
and computed SUSY algebra. The SUSY algebra with central charge
extension was obtained with a fixed choice of the operator ordering.

It would be interesting to check whether other choices of ordering
yield the same SUSY algebra and to extend the present formalism to
$\mathcal{N}=4$ Grassmannian model in 2+1 dimensions.

\acknowledgments We thank Jeong-Hyuck Park for early participation.
We are also grateful to Masato Arai, Dongsu Bak and Rabin Banerjee
for useful discussions. The work of PO and SS is supported by the
Science Research Center Program of the Korea Science and Engineering
Foundation through the Center for Quantum Spacetime(CQUeST) of
Sogang University with grant No. R11-2005-021 and by Korea Research
Foundation Grant (R05-2004-000-10682-0). The work of HN is the
result of research activities (Astrophysical Research Center for the
Structure and Evolution of the Cosmos (ARCSEC)) and grant No.
R01-2006-000-10965-0 from the Basic Research Program supported by
KOSEF.

\end{document}